\documentclass[reprint, superscriptaddress, nofootinbib]{revtex4-1}

\usepackage{amsmath,amsfonts,amssymb}
\usepackage{appendix}
\usepackage{color}
\usepackage{datetime}
\usepackage{graphicx}
\usepackage{dsfont}
\usepackage{array}
\usepackage[citecolor=blue]{hyperref}

\def\c{\chi}

\def\eps{\varepsilon}
\def\f{\frac}

\def\G{\Gamma}

\def\l{\left}

\def\la{\langle}
\def\ra{\rangle}

\def\mc{\mathcal}

\def\m{\mu}

\def\nn{\nonumber}

\def\p{\partial}

\def\r{\right}

\def\be{\begin{equation}}
\def\ee{\end{equation}}

\def\bea{\begin{eqnarray}}
\def\eea{\end{eqnarray}}

\def\ba{\begin{array}}
\def\ea{\end{array}}

\def\bc{\begin{center}}
\def\ec{\end{center}}

\def\bl{\begin{flushleft}}
\def\el{\end{flushleft}}

\def\br{\begin{flushright}}
\def\er{\end{flushright}}

\def\bi{\begin{itemize}}
\def\ei{\end{itemize}}

\def\bt{\begin{tabular}}
\def\et{\end{tabular}}

\newcolumntype{P}[1]{>{\centering\arraybackslash}p{#1}}
\newcolumntype{M}[1]{>{\centering\arraybackslash}m{#1}}

\newcommand{\mL}{\mathcal{L}}
\newcommand{\mO}{\mathcal{O}}

\begin{document}

\title{Feynman diagrams and the large charge expansion in \texorpdfstring{$3-\eps$ dimensions}{3-epsilon dimensions}}
\author{Gil Badel}
\email{gil.badel@epfl.ch}
\affiliation{Theoretical Particle Physics Laboratory (LPTP), Institute of Physics, EPFL, Lausanne, Switzerland}
\author{Gabriel Cuomo}
\email{gabriel.cuomo@epfl.ch}
\affiliation{Theoretical Particle Physics Laboratory (LPTP), Institute of Physics, EPFL, Lausanne, Switzerland}
\author{Alexander Monin}
\email{alexander.monin@unige.ch}
\affiliation{Theoretical Particle Physics Laboratory (LPTP), Institute of Physics, EPFL, Lausanne, Switzerland}
\affiliation{Department of Theoretical Physics, University of Geneva, \\
24 quai Ernest-Ansermet, 1211 Geneva, Switzerland}
\author{Riccardo Rattazzi}
\email{riccardo.rattazzi@epfl.ch}
\affiliation{Theoretical Particle Physics Laboratory (LPTP), Institute of Physics, EPFL, Lausanne, Switzerland}
\date{\today}

\begin{abstract}
In \href{https://arxiv.org/abs/1909.01269}{arXiv:1909.01269} it was shown that the scaling dimension of the lightest charge $n$ operator in the $U(1)$ model at the Wilson-Fisher fixed point in $D=4-\eps$ can be computed semiclassically for arbitrary values of $\lambda n$, where $\lambda$ is the perturbatively small fixed point coupling. Here we generalize this result to the fixed point of the $U(1)$ model in $3-\eps$ dimensions. The result interpolates continuously between diagrammatic calculations and the universal conformal superfluid regime for CFTs at large charge. In particular it reproduces the expectedly universal  $\mO(n^0)$ contribution to the scaling dimension of large charge operators in $3D$ CFTs.
\end{abstract}
\maketitle

\section{Introduction}

It is known that, even in a weakly coupled Quantum Field Theory (QFT), there exist situations where the  ordinary Feynman diagram expansion fails. An example is given by amplitudes with a sufficiently large number of legs. This instance received some attention in the 90's, in the study of  the production of a large number of massive bosons in high-energy scattering \cite{many1,many3,Rubakov:1995hq,Son:1995wz}\footnote{The claims made in  a recent revival \cite{Khoze:2017tjt} seem controversial. See indeed \cite{Monin:2018cbi} for a critical perspective.}. Multilegged amplitudes also occur in the correlators of operators carrying a large conserved internal charge, whose properties indeed defy perturbation theory for large enough charge.

Recently, it has been  shown that,  in conformal field theory (CFT),  large charge operators can generally be associated, via the state-operator correspondence, to a superfluid phase of the theory on the cylinder \cite{Hellerman:2015nra,Monin:2016jmo,Bern1,BootstrapLargeQ}. The corresponding CFT data are then \emph{universally} described, regardless of the details of the underlying CFT, by an \emph{effective}  field theory (EFT) for the  hydrodynamic Goldstone modes \cite{Nicolis_Zoology,Son:2002zn} of the superfluid. The systematic derivative and field expansion of the resulting EFT coincide with an expansion in inverse powers of the charge.

While the effective superfluid description should equally well apply to  strongly and weakly coupled theories, in the latter case it is also possible to work directly in the full theory, bypassing the EFT construction, or, in fact, deriving it. This was recently illustrated in \cite{Badel:2019oxl}, by focusing on the two-point function of the charge $n$ operator $\phi^n$ in the $U(1)$ invariant  Wilson-Fisher fixed point in $4-\eps$ dimensions. For arbitrary  $n$, the scaling dimension $\Delta_{\phi^n}$  was  computed semiclassically by expanding the path integral around a non-trivial trajectory.
 The result can be structured as a loop expansion in  the coupling $\lambda\propto\eps$ while treating 
  $\lambda n$ as a fixed parameter, playing a role similar to that of   the  't Hooft  
coupling in large $N$ gauge theories\footnote{A similar double expansion exists at large $N$ \cite{Bern7}.}.
The result encompasses the small charge regime ($\lambda n\ll 1$), where ordinary diagrammatic perturbation theory also applies, and the large charge regime ($\lambda n\gg 1$), described by a superfluid phase. Similar ideas were also shown to apply in the context of $\mc{N}=2$ superconformal theories \cite{Grassi:2019txd}, with the double expansion remarkably associated to a dual matrix model description.

In this paper we apply this methodology to compute the scaling dimension of $\phi^n$ in $ (\bar\phi \phi)^3$ at its  conformally invariant point in $3-\eps$ dimensions. The result follows the same pattern observed in $ (\bar\phi \phi)^2$ in $4-\eps$ dimensions. Besides confirming the generality of the method \cite{Badel:2019oxl}, the main interest of $ (\bar\phi \phi)^3$ in $D=3-\eps$ lies in the possibility of non-trivially comparing to the universal predictions of the large charge EFT of 3D CFT \cite{Hellerman:2015nra}. Indeed  the $\beta$ function of $ (\bar\phi \phi)^3$ arises only at 2-loops. At the 1-loop level the theory is therefore conformally invariant at $D=3$ for any value of $\lambda$. At this order, as $\lambda n$ is varied from small to large, our formulae non trivially interpolate between the prediction of standard Feynman diagram computations  and those of the universal superfluid description of large charge states in 3D CFT. In particular 
for $t\equiv\lambda n/\sqrt{3}\pi\gg 1$ our result for the scaling dimension takes the form:
\begin{eqnarray}\nn
\Delta_{\phi^n}= &t^{3/2}&\left[c_{3/2}+c_{1/2} t^{-1}+
\ldots\right]\\&+
t^0 &\left[d_0+d_{-1}t^{-1}+\ldots\right]\,.
\label{eq:PredictionDelta}
\end{eqnarray}
with  $c$'s and $d$'s having specific calculable values. This result nicely matches the universal predictions of the large charge EFT.  Within the general EFT construction the $c_k$'s are model dependent Wilson coefficients, but the
$d$'s are universally calculable effects associated to the 1-loop Casimir energy. Our result for the $d$'s should thus match the general theory, and they do. In particular we find
\begin{equation}\label{eq:n0}
d_0=-0.0937255(3)
\end{equation}
in agreement with \cite{Monin:2016jmo}.
The prediction of $d_{-1}$ is similarly matched, but the statement is less direct  as it involves a correlation with the subleading corrections to the dispersion relations of the Goldstone; we discuss the precise expression in section \ref{secRes}. Previously, eq.s \eqref{eq:PredictionDelta} and \eqref{eq:n0} were verified at large $N$ for monopole operators \cite{Anton}; the results of Monte-Carlo simulations for the $O(2)$ model at criticality are consistent with the expansion \eqref{eq:PredictionDelta} \cite{LargeQMonteCarlo1}, though their present precision is not sufficient to check the universal prediction for $d_0$. Our paper provides an alternative verification
where the large charge regime is continuously connected, as $\lambda n$ is varied, to diagrammatic perturbation theory.

\section{Lagrangian and conventions}

We consider the following $U(1)$ symmetric theory in $D=3-\eps$ dimensional euclidean space-time
\be
\mc L = \p \bar \phi \p \phi +\f{\lambda_0^2} {36} \l ( \bar \phi \phi\r )^3\,.
\label{eq:phi6Lagrangian}
\ee
Within this convention for the Lagrangian, one can easily realize that $\lambda_0$ is the loop counting parameter by rescaling $\phi\rightarrow\phi/\sqrt{\lambda_0}$ .
The renormalized coupling and field are defined as
\be
\phi = Z_\phi [\phi], ~~ \lambda_0 = M ^ {\eps} \lambda Z_\lambda\,,
\label{eq:bareRenormalized}
\ee
where $M$ denotes the sliding scale. The $\beta$-function is given by \cite{Pisarski:1982vz} 
\be\label{eq:beta}
\f{\p \lambda}{\p\log M}\equiv\beta(\lambda)= \lambda\left[-\eps + \frac{7\lambda^2}{48\pi^2}+\mO\l(\frac{\lambda^4}{(4\pi)^4}\r)\right]\,.
\ee
For $\eps\ll 1$, this implies the existence of an IR-stable fixed point at
\be\label{eq:fixedPt_lambda}
\frac{\lambda_*^2}{(4\pi)^2} = \f{3}{7}\eps+\mO\l(\eps^2\r)\,.
\ee

Notice that the $\beta$-function \eqref{eq:beta} starts at two-loop order at $\eps=0$. Hence the model is conformally invariant up to $\mO(\lambda)$ in exactly $D=3$. This observation will be important for what follows. The field wave-function renormalization starts at four loops and we shall always neglect it in the following.

\section{Anomalous dimension of large charge operators}

In this paper we focus on the calculation of the scaling dimension of the $U(1)$ charge $n$ operator $\phi^n$, focusing on the regime $n\gg 1$. In complete analogy with the $(\phi\bar\phi)^2$ case discussed in \cite{Badel:2019oxl}, the diagrammatic calculation for the anomalous dimension takes the form
\begin{equation}\label{eq:pertTheory}
\gamma_{\phi^n}=n\sum_{\ell=1}\lambda^\ell P_{\ell}(n),
\end{equation}
where $P_{\ell}$ is a polynomial of degree $\ell$ for $\ell\leq n$, and of degree $n$ for $\ell>n$. Thus, the loop order $\ell$ contribution  grows as $\lambda^{\ell} n^{\ell+1}$ for $\ell\leq n$, implying that the diagrammatic expansion breaks down for sufficiently large $\lambda n$. Re-organizing the series in \eqref{eq:pertTheory}, the scaling dimension can also be expanded as
\begin{equation}\label{eq:semiclassics}
\Delta_{\phi^n}=n\l(\frac{D}{2}-1\r)+\gamma_{\phi^n}=\sum_{\kappa=-1}\lambda^{\kappa}\Delta_\kappa(\lambda n).
\end{equation}
The main result of \cite{Badel:2019oxl} is that it is possible to compute the functions $\Delta_\kappa(\lambda n)$ for arbitrary $\lambda n$ via a
perturbative \emph{semiclassical calculation} around a non-trivial saddle; the result can  be organized as an expansion in $\lambda\ll 4\pi$ while treating $\lambda n$ as a fixed  parameter, closely analogous to the 't Hooft coupling of large $N$ theories. The goal of this paper is to compute the leading term and the first subleading correction in \eqref{eq:semiclassics}.

The scaling dimension \eqref{eq:semiclassics} is a physical (scheme-independent) quantity only at the fixed-point \eqref{eq:fixedPt_lambda}. However, in light of the observation at the end of the previous section, working at order $\mO(\lambda)$ we can take $\eps\rightarrow 0$ without affecting the conformal invariance of the theory\footnote{Dimensional regularization is still used in the intermediate steps.}. The leading order term $\Delta_{-1}(\lambda n)$ and the one-loop correction $\Delta_0(\lambda n)$ are hence scheme-independent for generic $\lambda$. 

\begin{figure}[t]
    \centering
    \includegraphics[width=3.5cm]{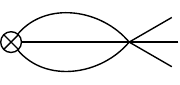}
    \caption{Two-loop diagram contributing to the $\phi^n$ anomalous dimension. The crossed circle denotes the $\phi^n$ insertion.\label{fig:gammaN_leading}}
\end{figure}

Working at fixed $n$, at leading order in $\lambda$, the anomalous dimension of $\phi^n(x)$ is determined by the diagram in  Fig.~\ref{fig:gammaN_leading} and it is given by 
\be
\gamma_{\phi^n} = \f{\lambda^2 n(n-1)(n-2)}{36(4\pi)^2}+\mO\l(\frac{\lambda^4 n^5}{(4\pi)^4}\r)\,.
\label{eq:perturbativeGamma}
\ee
Comparing with \eqref{eq:semiclassics}, one can readily extract the lowest order terms in the expansion of $\Delta_{-1}$ and $\Delta_0$ at small $\lambda n$.
We will use this expression as a check of the more general result that we will derive in the next section.

\section{Semiclassical computation}

To compute the scaling dimension $\Delta_{\phi^n}$ for arbitrary $\lambda n$ we proceed as in \cite{Badel:2019oxl}. Here we review the logic and outline the main steps. 

We first use a Weyl transformation to map the theory to the cylinder $\mathbb{R}\times S^{D-1}$. Parametrizing $\mathbb{R}^D$ by polar coordinates $(r,\Omega_{D-1})$, where  $\Omega_{D-1}$  collectively denotes the coordinates on $S^{D-1}$, and $\mathbb{R} \times S^{D-1}$ by $(\tau,\Omega_{D-1})$, the mapping is simply given by $r=Re^{\tau/R}$ with $R$ the sphere radius~\cite{Rychkov:2016iqz,Simmons-Duffin:2016gjk}. The Lagrangian of the theory on the cylinder reads:
\begin{equation}
\mL_{cyl}= \p \bar \phi \p \phi +m^2\bar{\phi}\phi+\f{\lambda^2} {36} \l ( \bar \phi \phi\r )^3\,,
\end{equation}
where $m^2=\l(\frac{D-2}{2R}\r)^2\overset{D=3}{=}\frac{1}{4R^2}$ arises from the ${\cal R}(g)\bar \phi\phi$ coupling to the Ricci scalar which is enforced by conformal invariance. Working at $\mO(\lambda)$,  we neglect the difference between bare and renormalized coupling, as that arises at $\mO(\lambda^2)$.

For small $\lambda n$, when diagrammatic perturbation theory holds, $\phi^n$ is the operator of lowest dimension with $U(1)$ charge $n$. Then for generic $\lambda n$,  
we \emph{define} the operator $\phi^n$ to be the lowest dimension charge $n$ operator. While this seems natural to us, the precise relation between such lowest dimension
operator and its explicit functional expression in terms of field variables  ($\phi^n$, $ \phi^{n-2}\partial^2 (\phi)^2$, etc.)
in the path integral, is not  obvious in the regime $\lambda n\gg 4\pi$. It should however become clear from our discussion that the precise form of the lowest dimension operator is a separate issue. It does not affect our computation of its scaling dimension but it matters for the computation of the normalization of the correlator, and thus for the computation of higher point functions.
We  plan to explore this in future work.
\footnote{In \cite{Badel:2019oxl} the analyticity of  $\Delta_{\phi^n}$ in $\lambda n$ as it directly emerges from the computation was taken as indication that there is no level crossing as $\lambda n$ is varied. However, unlike argued in \cite{Badel:2019oxl}, we now realize that does not imply that the field expression for the lowest dimension charge $n$ operator remains $\phi^n$ for all values of $\lambda n$.}.

According to the  above natural definition, $\Delta_{\phi^n}$ is directly determined by studying  the expectation value of the evolution operator $e^{-HT}$ in an arbitrary state $| \psi _n  \ra$ with fixed charge $n$ \footnote{Alternatively, one could include explicitly the operator insertions in the action as sources; this was done in \cite{Arias-Tamargo:2019xld} in the limit $\lambda^2 n^2\ll(4\pi)^2$.}. As long as there is an overlap between the state $| \psi _n \ra$ and the lowest energy state (with charge $n$), in the limit $T \to \infty$ the expectation gets saturated by the latter 
\be
\la \psi_n | e^{-HT} | \psi _n \ra \underset{T\to \infty}{=} \tilde{ \mc N} e^{-E_{\phi^n} T}\,,~~
E_{\phi^n}=\Delta_{\phi^n}/R\,.
\label{eq:Evolution}
\ee
To pick a specific state, we work in polar coordinates for the field:
\be
\phi = \f{\rho}{\sqrt{2}} e^{i \c}, ~~ \bar\phi = \f{\rho}{\sqrt{2}} e^{-i \c}\,.
\label{eq:waveFunction}
\ee
Following \cite{Monin:2016jmo}, we then consider the following path integral:
\begin{equation}\label{eq:PI1}
\begin{split}
\la \psi_n | e^{-HT} | \psi _n \ra = \mathcal{Z}^{-1}&\int \mc D \c_i \mc D \c_f 
\psi_n(\chi_i)\psi_n^*(\chi_f)\\
\times& \int^{\rho=f,~ \c=\c_f}_{\rho=f, ~\c=\c_i} \mc D \rho \mc D \c e^{-S}\,,
\end{split}
\end{equation}
where the insertions of the wave-functional
\begin{equation}
\psi_n(\chi)=\exp\l( \frac{i\,n}{R^{D-1}\Omega_{D-1}}\int d\Omega_{D-1}\c\r)
\end{equation}
ensure that the initial and final states have charge $n$, while the boundary value $f$ for $\rho$ is arbitrary and will be fixed later by convenience. The factor $\mathcal{Z}$ ensures that the vacuum to vacuum amplitude is normalized to unity:
\be
\mathcal{Z}=\int \mc D \rho \mc D \c e^{-S}\,.
\ee
The structure of the expansion \eqref{eq:semiclassics} follows from performing the path-integral in a saddle-point approximation. This is easily
seen rewriting eq. \eqref{eq:PI1} as
\be\label{eq:PI2}
\la \psi_n | e^{-HT} | \psi _n \ra =\mathcal{Z}^{-1}\int^{\rho=f}_{\rho=f} \mc D \rho \mc D \c e^{-S_{eff}}\,,
\ee
where the action on the right hand side is given by
\begin{multline}
S_{eff} =  \int_{-T/2}^{T/2} d\tau\int d\Omega_{D-1} \l [ \f{1}{2} (\p \rho)^2 +  \f{1}{2} \rho^2 ( \p \c ) ^ 2 \r.\\ 
\l.+ \f{m^2}{2} \rho^2 +
\f{\lambda^2}{2(12)^2} \rho^6 +i\frac{n}{R^{D-1}\Omega_{D-1}}\,\dot{\chi}\r ]\,.
\label{eq:effActCyl}
\end{multline}
Rescaling then the field as $\rho\rightarrow\rho/\lambda^{1/2}$ and collecting  up front the overall  $\lambda^{-1}$, one immediately recognizes eq. \eqref{eq:semiclassics} as the result of performing the path-integral \eqref{eq:PI2} as a systematic loop expansion around  a saddle-point (see \cite{Badel:2019oxl} for details).

Properly tuning the initial and final value $\rho_i=\rho_f=f$ in eq. \eqref{eq:PI2}, the stationary configuration for the action \eqref{eq:effActCyl} is given by a superfluid configuration:
\be
\rho = f\,, \qquad \c= -i \m \tau +\text{const.}\,,
\label{eq:cylinderSol}
\ee 
where $\mu$ is interpreted as the chemical potential of the system and $\mu$ and $f$ satisfy
\be\label{eq:EOMs}
\mu^2-m^2=\frac{\lambda^2}{48}f^4,\qquad
\mu f^2 R^{D-1}\Omega_{D-1}=n\,.
\ee
Given the constraint $f^2\geq 0$, the eq.s \eqref{eq:EOMs} admit a unique solution. In particular, in $D=3$ and for $n>0$,  $\mu$ reads:
\be\label{eq:mu}
R \m = \f{\sqrt{1+\sqrt{1+\frac{\lambda^2 n^2}{12\pi^2}}}}{2\sqrt{2}}\,.
\ee
For $\lambda n<0$ the chemical potential is given by minus the expression in \eqref{eq:mu} and is hence discontinuous at $\lambda n=0$ \footnote{This discontinuity is required as the scaling dimension of $\phi^n$ and the conjugated operator, $\bar{\phi}^n$, must be the same.}. In the following we always assume $\lambda n>0$. 

Plugging the solution into the classical action we extract the leading order contribution to the scaling dimension:
\begin{equation}
 S_{eff} /T=\frac{n }{3}\left(2\mu+\frac{m^2}{\mu}\right)\overset{D=3}{=}\frac1R\frac{\Delta_{-1}(\lambda n)}{\lambda}\,.
\end{equation}
Explicitly, the result reads
\begin{equation}\label{eq:DeltaMinus1}
\Delta_{-1}(\lambda n)=\lambda n\,F_{-1}\l(\frac{\lambda^2 n^2}{12\pi^2}\r)\,,
\end{equation}
where
\begin{equation}
F_{-1}(x)=\f{1+\sqrt{1+x}+x/3}{\sqrt{2}\,(1+\sqrt{1+x})^{3/2}}\,.
\end{equation}

To compute the one-loop correction we expand the fields around the saddle point configuration:
\begin{align}
\rho(x) = f +r(x)\,, \quad\c(x) = -i\mu \tau + \f{1}{f \sqrt{2}}\pi(x)\,.
\end{align}
The action \eqref{eq:effActCyl} at quadratic order in the fluctuations reads
\begin{multline}
S^{(2)} = \int^{T/2}_{-T/2} d\tau\int d\Omega_{D-1}\left[\f12(\p r)^2+\f12(\p \pi)^2 \r.\\ \l. \phantom{\f12(\p r)^2}
-2i\m\, r\p_\tau\pi +2(\m^2-m^2)r^2\r] \,.
\label{eq:cylAction2}
\end{multline}
This action describes two modes, with dispersion relations given by
\be
\omega^2_\pm (\ell) = J^2_\ell+2(2\m^2-m^2)\pm 2\sqrt{ J^2_\ell\m^2+(2\m^2-m^2)^2}\,,
\label{eq:dispersions}
\ee
where $J^2_\ell = \ell (\ell+D-2)/R^2$ is the eigenvalue of the Laplacian on the sphere. The first mode has a gap $\omega_+(0)=2\sqrt{2\mu^2-m^2}\propto \sqrt{\lambda n}$ for $\lambda n\gg 1$. The dispersion relation $\omega_-(\ell)$ describes instead a gapless mode, the Goldstone boson for the spontaneously broken $U(1)$ symmetry. These modes, except for the zero mode of the Goldstone which relates different charge sectors, provide a basis for the Fock space describing charge $n$ operators with higher scaling dimension. In particular,  the descendants, obtained by acting $q$ times with the $P_i$ generators of the conformal algebra, correspond to states involving a number $q$ of massless spin one quanta, each increasing the energy by $\omega_{-}(1)=1/R$.
Other modes describe different primaries; non-trivially, the expressions \eqref{eq:dispersions} include the leading $\lambda n$ corrections to the free theory values, effectively resumming the effect of infinitely many loop diagrams in standard diagrammatic calculations. 

In the large $\lambda n$ regime we can integrate out the gapped mode and describe the lightest states at charge $n$ through the superfluid effective theory for the gapless mode \cite{Monin:2016jmo}. In this limit the dispersion relation of the  Goldstone boson can be expanded in inverse powers of $\lambda n$ and reads
\begin{multline}\label{eq:Goldstone_dispersion}
R\omega_{-}(\ell)=\l[\frac{1}{\sqrt{2}}
-\frac{\sqrt{3} \pi }{\sqrt{2}\lambda  n }
+\mO\l(\frac{1}{(\lambda n)^2}\r)\r]J_\ell\\
+\left[\frac{\sqrt{3} \pi  }{2\sqrt{2}  }
+\mO\l(\frac{1}{\lambda n}\r)\r]
\frac{J_{\ell}^3}{\lambda n}
+\mO\l(\frac{J_{\ell}^5}{(\lambda n)^2}\r)\,.
\end{multline}
From this expression we see that the Goldstone sound speed approaches the value $c_s=1/\sqrt{2}$ as $\lambda n\rightarrow\infty$, as dictated by conformal invariance in the superfluid phase.

The one-loop correction $\Delta_0$ is determined by the fluctuation determinant around  the leading trajectory  \eqref{eq:cylinderSol}. Explicitly, we find\footnote{In \eqref{eq:one-loop-det1} we neglect the integration over the zero mode associated to the $U(1)$ symmetry, whose result is independent of $T$ and hence does not contribute to $E_{\phi^n}$ in eq. \eqref{eq:Evolution}.}
\be\label{eq:one-loop-det1}
\Delta_0(\lambda n)= \frac{1}{2}\sum_{\ell=0}^\infty n_{\ell}\left[\omega_+(\ell)+\omega_-(\ell)-2\omega_0(\ell)\right]\,, 
\ee
where $\omega_0^2(\ell)=
J_{\ell}^2+m^2=\left(\ell+\frac{D-2}{2}\right)^2/R^2$ is the free theory dispersion relation and $n_\ell=\f{(2\ell+D-2) \G(\ell+D-2)}{\G(\ell+1) \G(D-1)}$ is the multiplicity of the Laplacian on the $(D-1)$-dimensional sphere.
The analytic continuation to negative $D$ of the sum \eqref{eq:one-loop-det1} is convergent; the final result is finite in the limit $D\rightarrow 3$, consistently with the coupling not being renormalized at one-loop. Eventually, $\Delta_0$ can be written in terms of an infinite convergent sum as in \cite{Badel:2019oxl}
\be\label{eq:Delta0}
\Delta_0(\lambda n)=\frac{1}{4}-3 (R\mu)^2+\frac{\sqrt{8R^2\m ^2-1}}{2}+\frac12\sum_{\ell=1}^\infty\sigma(\ell)\,,
\ee
where $\sigma(\ell)$ is obtained from the summand in \eqref{eq:one-loop-det1} by subtracting the divergent piece:
\begin{multline}
\sigma(\ell)=(1+2\ell)R\l[\omega_+(\ell)+\omega_-(\ell)\r]\\-4\ell\l(\ell+1\r)-\l(6R^2\m^2-\frac{1}{2}\r)\,.
\end{multline}
In \eqref{eq:Delta0} all quantities are evaluated in $D=3$, hence $\m$ is given by eq. \eqref{eq:mu} and $m=\frac{1}{2R}$.

\section{Analysis of the result}\label{secRes}

Eq.s \eqref{eq:DeltaMinus1} and \eqref{eq:Delta0} provide the first two terms of the expansion \eqref{eq:semiclassics} for the scaling dimension of the operator $\phi^n$, $\Delta_{\phi^n}$. The result holds for arbitrary values of $\lambda n$. Here we explicitly show that $\Delta_{\phi^n}$ matches the diagrammatic result \eqref{eq:perturbativeGamma} and the large charge prediction \eqref{eq:PredictionDelta} in the two extreme regimes of, respectively, small and large $\lambda n$. 

Let us consider first the small $\lambda n$ regime. From eq. \eqref{eq:mu} it follows that the chemical potential, and consequently all the functions $\Delta_\kappa$, can be expanded in powers of $\lambda^2 n^2$. Explicitly neglecting terms of order $\mO\l(\frac{\lambda^6 n^7}{(4\pi)^6}\r)$, we get:
\begin{multline}\label{eq:smallLambdaN}
\Delta_{\phi^n}=\frac{n}{2}+\frac{\lambda^2}{(4\pi)^2}\l[\frac{n^3-3n^2}{36}+\mO\l(n\r)\r]\\
-\frac{\lambda^4}{(4\pi)^4}\l[
\frac{n^5}{144}-\frac{n^4(64-9\pi^2)}{1152}
+\mO\l(n^3\r)\r]+\ldots\,.
\end{multline}
In this regime we can compare eq. \eqref{eq:smallLambdaN} with the diagrammatic result \eqref{eq:perturbativeGamma}, finding perfect agreement.

Let us now discuss the large $\lambda n$ regime. The classical result \eqref{eq:DeltaMinus1} is easily seen to admit an expansion in inverse powers of $\lambda n$ with the expected form. The one-loop contribution \eqref{eq:Delta0} can be evaluated numerically for large $\mu\sim(\lambda n)^{1/2}$ and then fitted\footnote{We computed $\Delta_0$ numerically for $R\mu=10,11,\ldots 210$ to perform the fit; the final results are obtained using six fitting parameters in the expansion \eqref{eq:PredictionDelta}.} to the functional form \eqref{eq:PredictionDelta}. When doing this we also verified that the coefficients of terms which might modify the form of the expansion, such as a term linear in $\lambda n$, are compatible with zero within the numerical uncertainty.
The final result reads
\begin{eqnarray}\nn
\Delta_{\phi^n}= \,&t^{3/2}&\left[c_{3/2}+
c_{1/2}t^{-1}+
c_{-1/2}t^{-2}
+\ldots\right]\\
&+&\left[d_0\,+
d_{-1}
t^{-1}+\ldots\right]\,,
\label{eq:resultLarge}
\end{eqnarray}
where we defined $t=\frac{\lambda n}{\sqrt{3}\pi}$ and
the coefficients read
\begin{align}\nn
   &c_{3/2}=  \frac{\sqrt{3}\pi}{6\lambda}
-0.0653+\mO\left(\frac{\lambda}{\sqrt{3}\pi}\right)\,,\\ \nn
&c_{1/2}= \frac{\sqrt{3}\pi}{2\lambda}
+0.2088+\mO\left(\frac{\lambda}{\sqrt{3}\pi}\right)\,, \\ \label{eq:Q0}
&c_{-1/2}=  -\frac{\sqrt{3}\pi}{4\lambda}
-0.2627+\mO\left(\frac{\lambda}{\sqrt{3}\pi}\right) \,, \\ \nn
&d_0=-0.0937255(3)\,,\\ \nn
& d_{-1}=0.096(1)+\mO\l(\frac{\lambda}{\sqrt{3}\pi}\r)\,.
\end{align}
The parentheses show the numerical error on the last digit, when the latter is not negligible at the reported precision. 

To interpret this result notice that, as already mentioned above eq. \eqref{eq:Goldstone_dispersion}, in the large $\lambda n$ regime we can integrate out the gapped mode. We are then left with an \emph{effective theory} for the Goldstone boson on the cylinder. The form of the latter is determined by $U(1)$ and Weyl invariance and, in $D=3$, reads:
\begin{eqnarray}\nn
\mL/\sqrt{g} &= &-\frac{1}{\lambda}\left\{\alpha_1|\p\chi|^3+
\alpha_2\mathcal{R}_{\mu\nu}\frac{\p^\mu\chi\p^\nu\chi}{|\p\chi|}
\right.
\\
 &-&\left. \alpha_3|\p\chi|\left[ {\cal R}
+2\frac{\left(\p_\mu|\p\chi|\right)^2}{|\p\chi|^2}\right]+\ldots\right\}\,.\,
\label{eq:EFT}
\end{eqnarray}
The field is expanded around the classical value $\chi=-i\m \tau$ and the factor $1/\lambda$ in front ensures that the Wilson coefficients $\alpha_i$ scale as $ \lambda^0$.
In the EFT the derivative expansion coincides with an expansion in inverse powers of $\lambda n$; the loop counting parameter is $\lambda/(\lambda n)^{3/2}$ instead. It follows that the scaling dimension of the lightest charged operator takes the form \eqref{eq:resultLarge}, where the first line corresponds to short distance (classical plus quantum) contribution from both the radial and Goldstone mode, while the second line corresponds the one-loop Casimir energy of the Goldstone mode. This second contribution is thus a genuinely  long distance one. Matching the explicit calculation in the full theory with the result of the effective theory we can determine the Wilson coefficients $\alpha_1$ and $\alpha_3$ to next to leading order in $\lambda$ through the relations:
\begin{equation}\label{eq:alphas}
\lambda c_{3/2}=\frac{\pi }{3^{3/4} \sqrt{\alpha_1} },\qquad
\lambda c_{1/2}=\frac{4 \pi  \alpha_3 }{3^{1/4}\sqrt{\alpha_1  }}.
\end{equation}
From these we extract $\alpha_1=4/\sqrt{3}+0.3326 \,\lambda+\mO(\lambda^2)$ and $\alpha_3=\sqrt{3}/4+0.0644 \,\lambda+\mO(\lambda^2)$.
Notice that the coefficient $\alpha_2$ does not contribute to the scaling dimension at order $(\lambda n)^{1/2}$ since $\mathcal{R}_{00}=0$.

To discuss the value of the coefficients $d$'s in eq. \eqref{eq:Q0}, notice first that from the Lagrangian \eqref{eq:EFT} one derives the dispersion relation of the Goldstone boson as:
\begin{multline}\label{eq:Appendix2}
\hspace*{-5pt}
R\omega_-(\ell)=\l[\frac{1}{\sqrt{2}}-\frac{4\pi (\alpha_3+2 \alpha_2)}{\sqrt{2}\,\lambda n}+\mO\l(\frac{1}{(\lambda n)^2}\r)\r]J_\ell\\+\left[\sqrt{2}\pi
\alpha_3+\mO\l(\frac{1}{\lambda n}\r)\r]
\frac{J_{\ell}^3}{\lambda n}+\ldots\,.
\end{multline}
Comparing this equation to eqs. \eqref{eq:Goldstone_dispersion}, \eqref{eq:Q0} and \eqref{eq:alphas}, at leading order we find $\alpha_2=0$, and we can also check the consistency of the result for $\alpha_3$: $\alpha_3={\sqrt 3}/4$. \footnote{That $\alpha_2=0$ at the tree level in the effective lagrangian simply follows from the fact that, in the microscopic lagrangian, $\chi$ only appears through $(\partial \chi)^2$.}
Moreover with \eqref{eq:Appendix2} we can compute the one-loop Casimir energy of the Goldstone mode and determine the second line of \eqref{eq:resultLarge} in terms of the EFT Wilson coefficients:
\begin{align}\label{eq:PredictionQ0}
&d_0=-0.0937255\,,\\ \label{eq:PredictionQ1}
&d_{-1}=\alpha_2\times
0.4329 +\alpha_3\times 0.2236 
 \,.
\end{align}
As remarked in \cite{Hellerman:2015nra}, $d_0$ is a theory independent number.
The result of the explicit computation in the full model \eqref{eq:Q0} agrees with  its value \eqref{eq:PredictionQ0} almost to seven digits accuracy. Using the previously extracted values for the $\alpha_i$, the EFT prediction in eq. \eqref{eq:PredictionQ1} gives $d_{-1}=0.0968$, again in agreement with the explicit result in eq. \eqref{eq:Q0} within its  numerical accuracy.

\section{Conclusions}

In conclusion, in the tricritical $U(1)$  CFT in $3-\eps$ dimensions we computed the scaling dimension of the operator $\phi^n$ at the next-to-leading order  in the coupling $\lambda$, but for arbitrary values of $\lambda n$.
Our results nicely interpolate between  the small $\lambda n$ regime, when it is given by \eqref{eq:smallLambdaN}, in agreement with diagrammatic calculations, and  the large $\lambda n$ regime, where it reads as in \eqref{eq:resultLarge} and it agrees with the expectation for the universal conformal superfluid phase of CFTs at large charge. The remarkable agreement between the form of the quantum corrections in eq.s \eqref{eq:PredictionQ0} and \eqref{eq:PredictionQ1} and the explicit result \eqref{eq:resultLarge} provides a nontrivial check of the validity of our methodology.

By further developing these ideas, in the future it would be interesting to study the transition from diagrammatic perturbation theory to semiclassics in other observables studied by the large charge expansion in CFT. Possible examples include three-point functions of charged operators \cite{Monin:2016jmo} or the scaling dimension of charged operators with large spin \cite{vortices1,vortices2}. Perhaps, these ideas might be applied as well in the study of inhomogeneous phases, which are conjectured to describe operators in mixed symmetric representations of the $O(n)$ models \cite{HellermanO41,HellermanO42}.

\subsection*{Acknowledgments}

We would like to thank Anton de la Fuente, Zohar Komargodski, Joao Penedones and Slava Rychkov for useful discussions. The work of G.B., G.C. and R.R. is partially supported by the Swiss National Science Foundation under contract 200020-169696 and through the National Center of Competence in
Research SwissMAP. The work of A.M. is supported by ERC-AdG-2015 grant 694896. We acknowledge the Simons Collaboration on the Non-perturbative Bootstrap, KITP at U.C. Santa Barbara, Perimeter Institute, ICTP-SAIFR in Sao Paulo, Columbia University, the Simons Center for Geometry and Physics and the organizers of the workshop Quantum-Mechanical Systems at Large Quantum Number for hospitality and support during the completion of this work.

\bibliographystyle{JHEP}
\bibliography{epsilon}{}

\providecommand{\href}[2]{#2}\begingroup\raggedright\begin{thebibliography}{10}

\bibitem{many1}
L.~S. Brown, \emph{{Summing tree graphs at threshold}},
  \href{https://doi.org/10.1103/PhysRevD.46.R4125}{\emph{Phys. Rev.} {\bfseries
  D46} (1992) R4125} [\href{https://arxiv.org/abs/hep-ph/9209203}{{\ttfamily
  hep-ph/9209203}}].

\bibitem{many3}
M.~B. Voloshin, \emph{{Multiparticle amplitudes at zero energy and momentum in
  scalar theory}},
  \href{https://doi.org/10.1016/0550-3213(92)90678-5}{\emph{Nucl. Phys.}
  {\bfseries B383} (1992) 233}.

\bibitem{Rubakov:1995hq}
V.~A. Rubakov, \emph{{Nonperturbative aspects of multiparticle production}},
  in \emph{{2nd Rencontres du Vietnam: Consisting of 2 parallel conferences:
  Astrophysics Meeting: From the Sun and Beyond / Particle Physics Meeting:
  Physics at the Frontiers of the Standard Model, Ho Chi Minh City, Vietnam,
  October 21-28, 1995}}, 1995,
  \href{https://arxiv.org/abs/hep-ph/9511236}{{\ttfamily hep-ph/9511236}}.

\bibitem{Son:1995wz}
D.~T. Son, \emph{{Semiclassical approach for multiparticle production in scalar
  theories}}, \href{https://doi.org/10.1016/0550-3213(96)00386-0}{\emph{Nucl.
  Phys.} {\bfseries B477} (1996) 378}
  [\href{https://arxiv.org/abs/hep-ph/9505338}{{\ttfamily hep-ph/9505338}}].

\bibitem{Khoze:2017tjt}
V.~V. Khoze and M.~Spannowsky, \emph{{Higgsplosion: Solving the Hierarchy
  Problem via rapid decays of heavy states into multiple Higgs bosons}},
  \href{https://doi.org/10.1016/j.nuclphysb.2017.11.002}{\emph{Nucl. Phys.}
  {\bfseries B926} (2018) 95}
  [\href{https://arxiv.org/abs/1704.03447}{{\ttfamily 1704.03447}}].

\bibitem{Monin:2018cbi}
A.~Monin, \emph{{Inconsistencies of higgsplosion}},
  \href{https://arxiv.org/abs/1808.05810}{{\ttfamily 1808.05810}}.

\bibitem{Hellerman:2015nra}
S.~Hellerman, D.~Orlando, S.~Reffert and M.~Watanabe, \emph{{On the CFT
  Operator Spectrum at Large Global Charge}},
  \href{https://doi.org/10.1007/JHEP12(2015)071}{\emph{JHEP} {\bfseries 12}
  (2015) 071} [\href{https://arxiv.org/abs/1505.01537}{{\ttfamily
  1505.01537}}].

\bibitem{Monin:2016jmo}
A.~Monin, D.~Pirtskhalava, R.~Rattazzi and F.~K. Seibold, \emph{{Semiclassics,
  Goldstone Bosons and CFT data}},
  \href{https://doi.org/10.1007/JHEP06(2017)011}{\emph{JHEP} {\bfseries 06}
  (2017) 011} [\href{https://arxiv.org/abs/1611.02912}{{\ttfamily
  1611.02912}}].

\bibitem{Bern1}
L.~Alvarez-Gaume, O.~Loukas, D.~Orlando and S.~Reffert, \emph{{Compensating
  strong coupling with large charge}},
  \href{https://doi.org/10.1007/JHEP04(2017)059}{\emph{JHEP} {\bfseries 04}
  (2017) 059} [\href{https://arxiv.org/abs/1610.04495}{{\ttfamily
  1610.04495}}].

\bibitem{BootstrapLargeQ}
D.~Jafferis, B.~Mukhametzhanov and A.~Zhiboedov, \emph{{Conformal Bootstrap At
  Large Charge}}, \href{https://doi.org/10.1007/JHEP05(2018)043}{\emph{JHEP}
  {\bfseries 05} (2018) 043}
  [\href{https://arxiv.org/abs/1710.11161}{{\ttfamily 1710.11161}}].

\bibitem{Nicolis_Zoology}
A.~Nicolis, R.~Penco, F.~Piazza and R.~Rattazzi, \emph{{Zoology of condensed
  matter: Framids, ordinary stuff, extra-ordinary stuff}},
  \href{https://doi.org/10.1007/JHEP06(2015)155}{\emph{JHEP} {\bfseries 06}
  (2015) 155} [\href{https://arxiv.org/abs/1501.03845}{{\ttfamily
  1501.03845}}].

\bibitem{Son:2002zn}
D.~T. Son, \emph{{Low-energy quantum effective action for relativistic
  superfluids}},  \href{https://arxiv.org/abs/hep-ph/0204199}{{\ttfamily
  hep-ph/0204199}}.

\bibitem{Badel:2019oxl}
G.~Badel, G.~Cuomo, A.~Monin and R.~Rattazzi, \emph{{The Epsilon Expansion
  Meets Semiclassics}},  \href{https://arxiv.org/abs/1909.01269}{{\ttfamily
  1909.01269}}.

\bibitem{Bern7}
L.~Alvarez-Gaume, D.~Orlando and S.~Reffert, \emph{{Large charge at large N}},
  \href{https://arxiv.org/abs/1909.02571}{{\ttfamily 1909.02571}}.

\bibitem{Grassi:2019txd}
A.~Grassi, Z.~Komargodski and L.~Tizzano, \emph{{Extremal Correlators and
  Random Matrix Theory}},  \href{https://arxiv.org/abs/1908.10306}{{\ttfamily
  1908.10306}}.

\bibitem{Anton}
A.~De~La~Fuente, \emph{{The large charge expansion at large $N$}},
  \href{https://doi.org/10.1007/JHEP08(2018)041}{\emph{JHEP} {\bfseries 08}
  (2018) 041} [\href{https://arxiv.org/abs/1805.00501}{{\ttfamily
  1805.00501}}].

\bibitem{LargeQMonteCarlo1}
D.~Banerjee, S.~Chandrasekharan and D.~Orlando, \emph{{Conformal dimensions via
  large charge expansion}},
  \href{https://doi.org/10.1103/PhysRevLett.120.061603}{\emph{Phys. Rev. Lett.}
  {\bfseries 120} (2018) 061603}
  [\href{https://arxiv.org/abs/1707.00711}{{\ttfamily 1707.00711}}].

\bibitem{Pisarski:1982vz}
R.~D. Pisarski, \emph{{Fixed point structure of $\phi^6$ in three-dimensions at
  large N}}, \href{https://doi.org/10.1103/PhysRevLett.48.574}{\emph{Phys. Rev.
  Lett.} {\bfseries 48} (1982) 574}.

\bibitem{Rychkov:2016iqz}
S.~Rychkov, \emph{{EPFL Lectures on Conformal Field Theory in $D \geq 3$
  Dimensions}}, SpringerBriefs in Physics. 2016,
  \href{https://doi.org/10.1007/978-3-319-43626-5}{10.1007/978-3-319-43626-5},
  [\href{https://arxiv.org/abs/1601.05000}{{\ttfamily 1601.05000}}].

\bibitem{Simmons-Duffin:2016gjk}
D.~Simmons-Duffin, \emph{{The Conformal Bootstrap}},  in \emph{{Proceedings,
  Theoretical Advanced Study Institute in Elementary Particle Physics: New
  Frontiers in Fields and Strings (TASI 2015): Boulder, CO, USA, June 1-26,
  2015}}, pp.~1--74, 2017, \href{https://arxiv.org/abs/1602.07982}{{\ttfamily
  1602.07982}}, \href{https://doi.org/10.1142/9789813149441_0001}{DOI}.

\bibitem{Arias-Tamargo:2019xld}
G.~Arias-Tamargo, D.~Rodriguez-Gomez and J.~G. Russo, \emph{{The large charge
  limit of scalar field theories and the Wilson-Fisher fixed point at
  $\epsilon=0$}},  \href{https://arxiv.org/abs/1908.11347}{{\ttfamily
  1908.11347}}.

\bibitem{vortices1}
G.~Cuomo, A.~de~la Fuente, A.~Monin, D.~Pirtskhalava and R.~Rattazzi,
  \emph{{Rotating superfluids and spinning charged operators in conformal field
  theory}}, \href{https://doi.org/10.1103/PhysRevD.97.045012}{\emph{Phys. Rev.}
  {\bfseries D97} (2018) 045012}
  [\href{https://arxiv.org/abs/1711.02108}{{\ttfamily 1711.02108}}].

\bibitem{vortices2}
G.~Cuomo, \emph{{Superfluids, vortices and spinning charged operators in 4d
  CFT}},  \href{https://arxiv.org/abs/1906.07283}{{\ttfamily 1906.07283}}.

\bibitem{HellermanO41}
S.~Hellerman, N.~Kobayashi, S.~Maeda and M.~Watanabe, \emph{{A Note on
  Inhomogeneous Ground States at Large Global Charge}},
  \href{https://doi.org/10.1007/JHEP10(2019)038}{\emph{JHEP} {\bfseries 10}
  (2019) 038} [\href{https://arxiv.org/abs/1705.05825}{{\ttfamily
  1705.05825}}].

\bibitem{HellermanO42}
S.~Hellerman, N.~Kobayashi, S.~Maeda and M.~Watanabe, \emph{{Observables in
  Inhomogeneous Ground States at Large Global Charge}},
  \href{https://arxiv.org/abs/1804.06495}{{\ttfamily 1804.06495}}.

\end{thebibliography}\endgroup

\end{document}